# Momentum-resolved surface enhanced Raman scattering from a nanowire-nanoparticle junction cavity

*Adarsh B. Vasista, Shailendra K. Chaubey, David J. Gosztola, Gary P. Wiederrecht, Stephen K. Gray and G. V. Pavan Kumar*\*

Dr. Adarsh B. Vasista, Shailendra K. Chaubey, and Prof. Dr. G. V. Pavan Kumar
Department of Physics, Indian Institute of Science Education and Research, Pune-411008, India
E-mail: pavan@iiserpune.ac.in

Dr. David J. Gosztola, Dr. Gary P. Wiederrecht, and Dr. Stephen K. Gray
Center for Nanoscale Materials, Argonne National Laboratory, Lemont, Illinois 60439, U.S.A

Prof. Dr. G. V. Pavan Kumar
Center for Energy Science, Indian Institute of Science Education and Research, Pune-411008, India

**Abstract**

Herein we report experimental evidence of directional SERS from molecules situated inside a single nanowire – nanoparticle junction cavity. The emission was confined to a narrow range of wavevectors perpendicular to the axis of the cavity. In addition to this, the molecules excite multiple guided modes of the nanowire which were imaged using leakage radiation Fourier microscopy. We further characterize the emission wavevectors as a function of output polarization. The excited guided modes of the wire show interesting polarization signatures. All the results were corroborated using finite element method based numerical simulations. Essentially, we provide an important connection between gap-cavity enhanced Raman scattering and its directionality of emission. The results may be of relevance in understanding the cavity electrodynamics at the nanoscale and molecular coupling to extremely small gaps between a one dimensional and a zero dimensional plasmonic nanostructure.

Cavity assisted molecular emission finds relevance in addressing various aspects of fundamental and applied scientific problems.[1-3] The study of nanocavities is important as they form quantum-mechanical test beds,[4] single molecule sensors,[5, 6] optical antennas,[7] systems providing large purcell enhancements,[8] etc. Due to a very large electromagnetic field enhancement and the modification of the local density of optical states, molecular emission inside extremely small cavities can be enhanced many fold. Nanocavities have been utilized to achieve single molecule strong coupling,[9] differentiating molecular fluorescence and Raman signals,[10] enhancing metal/molecular photoluminescence,[11, 12] etc. Hence studying nanocavities can potentially have wide importance not only in understanding fundamental cavity-electrodynamics but also in various applications.

Nanocavity assisted surface enhanced Raman scattering (SERS) has been studied, in the recent past, with an emphasis on intensity enhancement and sensing capabilities.[13, 14] An important aspect of emission which has not been studied extensively as far as cavity assisted SERS is concerned is the wavevector of emission. Projecting Raman signals onto the wavevector space can give extensive information about the molecular orientation,[15, 16] and molecular interaction with its environment.[17, 18]

A variety of cavity architectures have been utilized to understand molecule – cavity interaction including nanoparticle dimer cavity,[9, 19] particle on a mirror cavity,[9, 20] nanowire on a mirror cavity [8, 10, 21] etc. Nanoparticle dimers show interesting interaction between two quasi – zero dimensional structures. Nanoparticle and nanowire on mirror cavities, on the other hand, are used to study interaction between a quasi – zero and quasi – one dimensional structures with thin films, respectively. Nanowire on mirror shows very interesting polarization signatures and has reported to show differential wavevector distribution for SERS and fluorescence.[10] A complementary junction cavity along these lines is a nanowire – nanoparticle junction cavity. It provides a platform to probe the interaction between a quasi zero dimensional and a quasi one dimensional stucture.

With the above motivation in mind, we probed a novel nanocavity prepared using a nanoparticle and a nanowire. Single nanowire – nanoparticle junction cavities have unique advantages as they can be excited remotely. They are relatively easier to prepare using wet chemical approaches and do not require sophisticated lithography techniques. Molecules can be situated inside the nanowire – nanoparticle junction cavity and can be excited using an objective lens. Since the distance between nanoparticle and nanowire in a typical junction is on the order of 4 – 5 nm, the molecules will experience very large electric fields and hence enhanced Raman scattering. Probing momentum signatures of molecular SERS from these extremely small junctions can be very illuminating.

In the past, nanowire-nanoparticle junctions have been utilized to understand the input polarization dependence and SERS enhancement,[22-26] to enhance performance of opto-electronic devices,[27] directional elastic scattering and fluorescence,[28] remote-excitation SERS,[29, 30] to understand coupling,[31] etc. In all these experiments, either the molecular signature is not exactly from the junction alone or the momentum resolved imaging is not carried out. Also, previous reports on FP imaging on SERS are more oriented towards electronically resonant molecules and lithographically prepared samples. Here, we focus on a self-assembled structure, a nanowire – nanoparticle system, with good control over the placement of the molecule. Since the structure is chemically prepared and self-assembled, the structure will suffer less optical scattering and Ohmic losses from the surface defects and suits plasmon propagation related experiments.This calls for momentum resolved SERS measurements from molecular SERS emission only from the nanowire – nanoparticle junction cavity.

Here we present momentum space imaging of bi-phenyl thiol (BPT) molecular SERS by situating BPT molecules inside a single nanowire-nanoparticle junction. Using Fourier plane (FP) imaging, we show that a nanowire- nanoparticle cavity not only enhances the SERS emission, but also directs it to a specific range of wavevectors. Further we show that the SERS carries interesting output polarization signatures. We perform 3D finite element method (FEM) based numerical simulations to corroborate experimental findings.

Figure 1 shows a conceptual schematic of the experimental configuration. Silver nanowires of diameter~ 300 nm were prepared using a poly-ol process reported elsewhere.[32]

Gold nanoparticles of diameter ~ 200 nm (Sigma-Aldrich) were washed in acetone three times to reduce the surfactant coating and then soaked in a solution of BPT molecules overnight. Since the thiol based molecules have an affinity towards the surface of gold, they preferentially attach to the gold nanoparticle surface. Excess molecules were then washed off and the particles were redistributed in ethanol. Typical nanowire-nanoparticle junctions were prepared by dropcasting silver nanowires on a glass coverslip and allowed to dry. The nanoparticle solution was then dropcast onto the nanowire coated glass coverslip to create nanowire-nanoparticle junctions. The detailed procedure can be found elsewhere[33]. As the diameter of the nanowire is ~300 nm with pentagonal crossection and that of particle is 200 nm, the probability of the nanoparticle assemmbling on top of the nanowire is minimal. The particle usually assembles by the side of the nanowire, thus making the configuration repeatable. Silver nanowires are easy to synthesize and known to show better surface plasmon propagation properties. This makes it to be the preferred choice for undertaking plasmon propagation measurements. Gold nanospheres have plasmon resonances red shifted compared to the silver spheres. For the laser wavelength (632.8 nm available) gold spheres of size about 100 nm are a nearly perfect fit as they shows considerable extinction cross section and the resonance is near 630 nm. Hence we chose silver nanowire and gold nanospheres for our experimental work. Individual junctions were then excited using a high numerical aperture objective lens, 100x 1.49 NA oil immersion, and the backscattered light was collected through the same lens. The wavelength of illumination was 633 nm. The laser line was filtered, in the collection path, using a combination of notch and edge filters and the resulting light was projected onto an EM-CCD or to the spectrometer and analyzed (see figure S1 of the supplementary information for detailed experimental setup). Inset (i) of figure 1 shows a scanning electron microscopy (SEM) image of an individual nanowire-nanoparticle junction. Spectral information collected from the system is shown in inset (ii) of figure 1, containing sharp Raman modes of the BPT molecules over a broad inelastic background. The intensity of the laser line is minimal when compared to the Raman signals, ensuring that the captured Fourier plane (FP) images have a minimal elastically scattered component. It is important to note that the molecules in the system are non-resonant and have minimal fluorescing properties.

Molecules that are sandwiched in between the nanoparticle and the nanowire experience large electric field enhancement in the cavity. Since the molecules are anchored on the particle, the SERS signal is essentially generated at the junction (see figure S2 of the supplementary information). This location specificity provides an important handle towards cavity assisted SERS.

To understand the polarization characteristics of the system, we performed both input and output polarization resolved spectroscopy on the nanowire-nanoparticle junction. The junction was excited using two orthogonal input polarizations (transverse and longitudinal with respect to the axis of the wire) and the resulting SERS signals from the system were captured. The outcoupled light was collected from the entire nanowire-nanoparticle system and no spatial filtering was performed. Figure 2 (a) shows the input polarization dependence of the BPT Raman signals. It is clear that the SERS modes of the molecule are strong for transverse input polarization. This is due to excitation of a local gap plasmon mode between nanowire and nanoparticle.[24] For output polarization resolved measurements, we excite the junction with input polarization transverse to the axis of the wire (as we get maximum SERS intensity with transverse input polarization) and the resulting SERS signal was analyzed for the two orthogonal components. Figure 2 (b) shows a typical output polarized SERS spectra from the system. For output polarization transverse to the axis of the nanowire, we find a large SERS intensity when compared to longitudinal output polarization. Note that figure 2 (a) and (b) are spectra acquired from different junctions. For any polarization dependent measurement, the input intensity is kept constant.

After analyzing the spectral characteristics of the system, we now perform Fourier plane imaging of the outcoupled emission from the system, as shown in figure 3. Figure 3 (a) shows the real plane brightfield image of a nanowire-nanoparticle system. The junction was excited using a focused 633 nm laser beam with polarization transverse to the nanowire axis. The inset in Figure 3 (a) shows a zoomed SEM image of the junction clearly showing the geometry. This ensures that the system being probed is indeed a single nanoparticle – nanowire junction. The light outcoupled from the system was then captured and projected onto the Fourier plane. Figure 3 (b) shows the corresponding FP image after rejecting the laser line. The FP image has two important features: (i) Directional emission of the SERS signal perpendicular to the direction of the nanowire. (ii) Guided modes of the nanowire. The molecular SERS generated at the junction can either outcouple directly to free-space after getting modulated by the presence of the wire - particle junction or get coupled to the guided modes of the nanowire. Since the nanowire used here has a diameter of ~ 300 nm, it supports both leaky and bound modes.[34, 35] The SERS emission acts as a localized source for excitation of the guided surface plasmon modes of the wire. This enables the molecular SERS to outcouple in two different modes. To further establish this fact, we studied SERS emission from a nanowire of thickness ~ 100 nm. For thinner nanowires (size ~ 100 nm), the leaky modes cease to exist. The FP image for such a system should not contain the leaky guided modes. This is shown in figure S3 of the supplementary information. The emission directionality can also be measured by collecting the SERS signals outcoupled towards the air side of the sample as shown in figure S4 of supplementary information. The outcoupled SERS on the air side of the sample also shows a directional emission pattern. However since the majority of emission leaks to the glass side of the sample, larger enhancements are found when one collects SERS through the glass-side.

To quantify the emission further, we consider the line profile of emission across the emission maxima, as shown in figure 3 (b). Figure 3 (c) is the line profile drawn perpendicular to the emission of leaky guided modes in the FP image (red dash-dot line). The emission pattern shows two recognizable leaky modes of the wire emitting at $k/k_0$ = -1.04; 0.99 and -1.15; 1.13. As the nanowires are thick (~ 300 nm in diameter), one can see multiple leaky modes, at the Raman wavelength of the molecule. Next we quantify the emission from the junction. The profile of the black dotted line parallel to the leaky modes (see figure 3 (b)) is shown in figure 3 (d). The maximum SERS emission happens here at $k/k_0 = 1.02$, near the critical angle of the air-glass interface. The emission is also very sharp with the full width at half maxima (FWHM) $\Delta(k/k_0) = 0.21$. The system under consideration can be understood by the concept of a Yagi-Uda antenna, though the system is not designed according to the concepts of the Yagi-Uda antenna (a molecular dipole place between a director and a reflector). We may consider the nanowire and the nanoparticles as the reflector – director pair in this system. The emission profile of the molecular dipole is modified due to the presence of a reflector – director pair and the emission dipolar lobes are modified by making the emission biased towards one of the directions along the nanowire – nanoparticle junction.

With the understanding of the Fourier plane signatures of SERS emission from the nanowire-nanoparticle system, we measured output polarization resolved FP images. Polarization resolved imaging is very important for this system as the modes of the nanowire show rich polarization signatures and that can affect polarization of the SERS emission from the entire nanowire-nanoparticle system. Figure 4 shows the output polarization resolved Fourier plane images from an individual nanowire-nanoparticle system. Figure 4 (a) shows the real plane brightfield image of the system under study. The junction was excited using input polarizations longitudinal and transverse to the axis of the wire. The outcoupled SERS emission was analyzed for orthogonal polarization states (transverse and longitudinal to the nanowire axis). The output polarization resolved spectra is already shown in figure 2 (b). Figure 4 (b) shows unresolved SERS FP image from the system showing directional SERS emission from the junction and multiple leaky modes of the wire. Figure 4 (c) and (d) show output polarization resolved FP images for longitudinal and transverse output polarizations respectively. Both the images show directional emission from the junction and leaky modes. Wires which support multiple leaky modes have very rich polarization structure for the guided modes.[36] Guided modes can possess polarizations perpendicular and along the axis of the nanowire. In our case, the molecular dipoles couple to multiple guided modes of the nanowire. Thus we see guided modes for both of the output polarization states.

To further understand the process of emission from the nanowire – nanoparticle junction, we performed 3D finite element method (FEM) electrodynamic simulations using COMSOL 5.3. The silver nanowire was modeled with a pentagonal cross-section of diameter 300 nm and of length 10 µm. The nanoparticle is placed at the center of the nanowire (equidistant from the ends). The gold nanoparticle was modeled as a sphere of diameter 200 nm and placed close to the nanowire. The refractive indices of silver and gold were taken from Ref. [37]. The junction was illuminated with a focused Gaussian wave with incident wavelength 633 nm. The spot size was fixed to half of the illumination wavelength. The polarization of the illumination was kept transverse to the axis of the wire. To understand the coupling of SERS emission to the plasmon modes of the wire, an oscillating *xyz*- dipolar source at wavelength 703 nm (the wavelength at which the prominent Raman mode exists) was placed inside the cavity between nanowire and nanoparticle. The near field electric field captured was then projected onto the far field by using reciprocity arguments.[38]

Figure 5 (a) shows the cross-section near field electric field distribution of the system. One can see strong electric field enhancement at the junction between the nanoparticle and the nanowire. Hence the molecules sandwiched between the wire and the particle experience strong Raman enhancement (of the order $10^4$)(see figure S5 of the supplementary information for far field response of the structure at the Raman wavelength). Figure 5 (b) shows the far field distribution of scattering from the nanowire – nanoparticle junction. The far field scattering from the system is biased along $-k_x/k_0$, perpendicular to the axis of the wire. This qualitatively agrees with the experimental results, where the emission from the junction is biased towards the direction perpendicular to the axis of the nanowire (see Figure 3 (b)). However Figure 5 (a) and (b) do not provide information about the coupling of SERS to the plasmon modes of the wire.

To further model the near-field coupling of molecular SERS to the propagating plasmon modes of the wire, we placed three spatially co-incident oscillating point dipoles each oriented along x, y, and z axes respectively inside the nanowire – nanoparticle cavity. The dipolar emission now couples to the propagating modes of the wire as shown in the near field electric field map in Figure 5 (c). The propagating modes leak into the medium of higher refractive index as shown in Figure 5 (d). The far field image shows guided modes of the wire. This clearly shows that molecular emission from the cavity essentially generates the guided plasmon modes in the wire, which leak into the medium of higher refractive index. The emission from the junction itself follows the scattering pattern from the junction. Hence, the convolution of these factors affect the complete FP as seen in figure 3 (b). It should also be noted that there appears to be a slight directional anisotropy for guided modes of the nanowire (see figure 3 (b) and figure 5 (d) ). This may be due to the presence of the nanoparticle and should depend on the orientation and coupling of the molecular SERS to the guided modes of the geometery.

Exact and faithful modeling of SERS emission from a nanowire – nanparticle cavity is, of course, very challenging. Nevertheless this simple model of scattering and near-field excitation sheds light on the generation and directionality of the emission from the system under consideration.

In summary, we have shown that the single nanowire – nanoparticle junction cavity directs the SERS emission from molecules fixed inside the cavity to a narrow range of wavevectors. We also probed output polarization resolved momentum signatures from the system. The molecules inside the cavity also generate guided modes of the nanowire which have unique polarization properties. The experimental results are corroborated with FEM based numerical simulations. We believe that the results presented will have wider relevance in understanding molecular coupling to extremely small cavities, those prepared with a combination of a zero and one dimensional nanostrutures. Specifically in understanding coupling of SERS to the multiple guided modes of the geometry which shows very interesting and non-trivial momentum and polarization signatures. This experiment can be extrapolated to study single molecule SERS in remote excitation configurations, and can be harnessed as a potential test-bed for spin-orbit interaction based Raman emission.[39]

**Acknowledgments:**

This work was partially funded by Indo-French Centre for the Promotion of Advanced Research (IFCPAR) (55043), INSA grant, Center for Energy Science grant (SR/NM/TP-13/2016) and US Air Force Research Laboratory grant. G.V.P.K., G.P.W., and S.K.G. thank IUSSTF grant, Quantum Plasmonics of Hybrid Nano-Assembles/JC-3-2014. A.B.V thanks fruitful discussions with Dr. Rohit Chikkaraddy. A.B.V acknowldeges help from Ms. Debashree for preparing samples. Theory and analyis were performed, in part, at the Center for Nanoscale Materials, a U.S. Department of Energy Office of Science User Facility, and supported by the U.S. Department of Energy, Office of Science, under Contract No. DE-AC02-06CH11357.

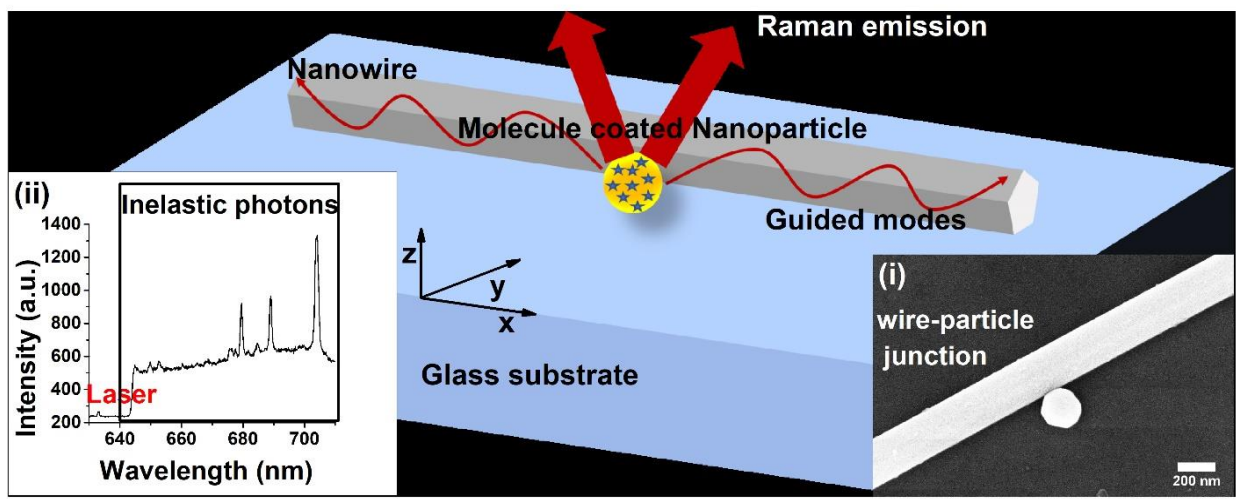

Figure 1: Schematic representing the experimental configuration. An individual silver nanowire – bi-phenyl thiol molecule coated gold nanoparticle junction was excited using a high numerical aperture objective lens. The molecular Raman signals will outcouple from the junction as directional photons and also couple to the guided modes of the wire. *Inset* (i) SEM image of the nanowire-nanoparticle junction. (ii) Spectra of the collected photons from the nanowire-nanoparticle system showing sharp Raman modes over inelastic background.

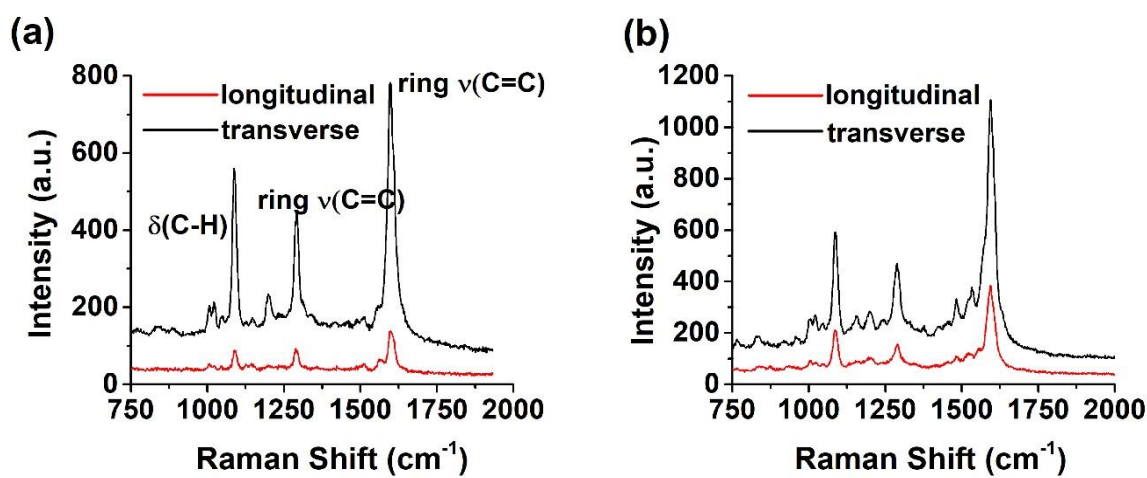

Figure 2: Polarization resolved Raman spectra. (a) Spectra collected from the nanowire-nanoparticle system for two orthogonal input polarizations. The Raman modes are intense for input polarization transverse to the axis of nanowire. (b) Output polarization resolved Raman

spectra. The input polarization was kept transverse to the axis of the wire. The Raman modes are intense for output polarization transverse to the axis of the wire.

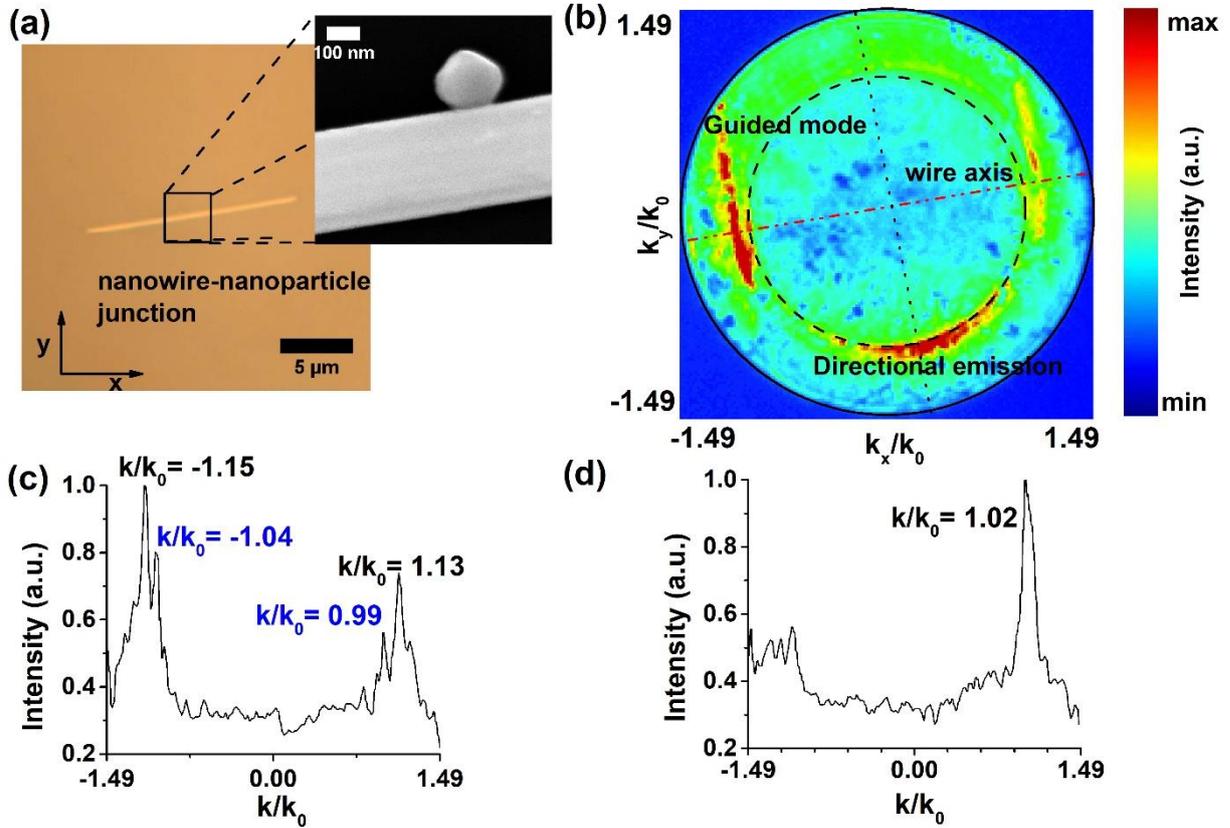

Figure 3: Fourier plane SERS imaging. (a) Real plane brightfield image showing a nanowire-nanoparticle junction. Inset shows the corresponding SEM image clearly showing the zoomed view of the junction. (b) Corresponding Fourier plane image captured by rejecting the input laser line. The SERS emission from the junction is directed to a narrow range of k-vectors. Also the SERS emission is coupled to the guided modes of the wire. (c) Line cut along the dash-dot line (red) in the FP image showing the wavevector of emission for guided modes. (d) Line cut along the dotted line (black) in the FP image showing the wavevector of emission for Raman modes from the junction.

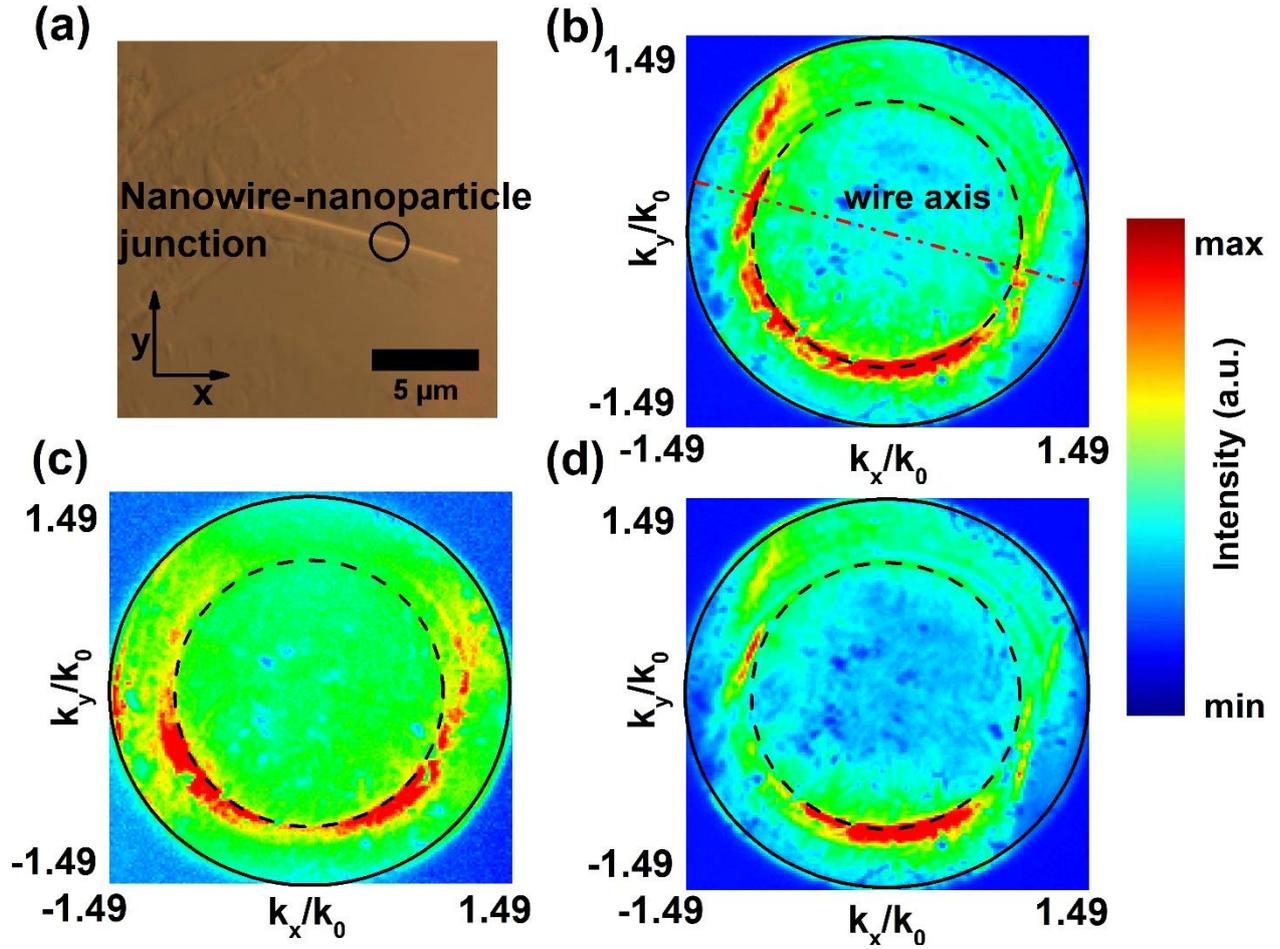

Figure 4: Output polarized Fourier plane Raman imaging. (a) Real plane brighfield image showing nanowire-nanoparticle junction. (b) Corresponding Fourier plane image captured by rejecting input laser line. The input polarization is kept transverse to the axis of wire. Output polarized FP images for polarization (c) longitudinal and (d) transverse with respect to the axis of the nanowire respectively.

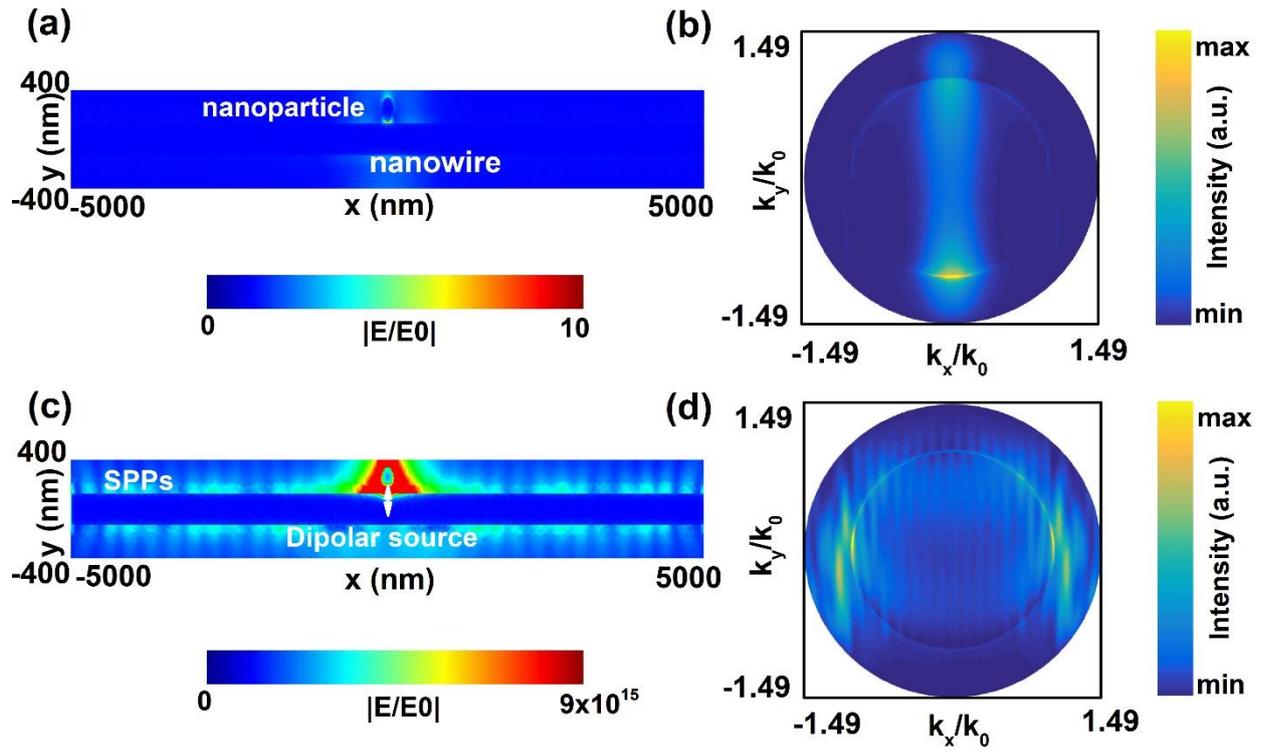

Figure 5: Numerical simulations. (a) Cross-sectional near field electric field map for far field Gaussian illumination at 633 nm. (b) Calculated FP pattern of scattering from the junction showing directional emission at 633 nm. (c) Cross-sectional near field electric field map for near field dipolar excitation at 703 nm. (d) Calculated FP pattern for dipolar excitation at 703 nm.